\documentclass{article}
\usepackage{lipsum} 
\usepackage{titling} 
\usepackage{amsmath}
\usepackage{amsfonts}
\usepackage{amssymb}
\usepackage{amsthm}          
\usepackage{natbib}
\usepackage{graphicx}
\usepackage{todonotes} 
\usepackage{tcolorbox} 
\usepackage{subcaption}
\usepackage{float}
\usepackage{verbatim}
\usepackage{url}
\usepackage{sectsty}
\usepackage{booktabs}
\usepackage{bm}
\usepackage{algorithm}     
\usepackage{algpseudocode}  
\usepackage[toc,page]{appendix}
\algrenewcommand\algorithmicrequire{\textbf{Inputs:}}  
\algrenewcommand\algorithmicensure {\textbf{Output:}} 
\sectionfont{\large\bfseries}
\newcommand{\ra}{\raisebox{.7ex}{$\rightarrow$}}
\tcbset{mylabel/.style={
  boxrule=0.4pt,
  colframe=black,
  colback=gray!15,
  arc=2pt,
  boxsep=1pt,
  left=2pt,right=2pt, top=1pt,bottom=1pt}}

\newtheorem{theorem}{Theorem}[section]
\newtheorem{corollary}[theorem]{Corollary}

\newcommand{\vect}[1]{\bm{#1}}

\renewenvironment{abstract}{
  \normalfont
  \begin{center}
    \large\bfseries Abstract
  \end{center}
}{}

\title{%
  \rule{\linewidth}{1pt} \\[1ex]
  \textbf{Distributed Training under Packet Loss} \\[0.5ex]
  \rule{\linewidth}{0.5pt}
}

\author{
  Erez Weintraub, Ron Banner, Ariel Orda
}

\date{June, 2025} 

\begin{document}

\maketitle

\begin{abstract}
\textbf{Background.}
State-of-the-art language and vision models are routinely trained across thousands of GPUs, often spanning multiple data-centers, yet today’s distributed frameworks still assume reliable connections (e.g., InfiniBand or RoCE). The resulting acknowledgment traffic and retransmissions inflate tail latencies and limit scalability. Leveraging unreliable connections will reduce latency but may sacrifice model's accuracy and convergence once packets are dropped. A principled, end-to-end solution that preserves accuracy and convergence guarantees under genuine packet loss has previously been missing.\\
\textbf{Results.}
We address this critical gap by introducing a novel distributed training framework capable of operating over unreliable connections, offering unbiased gradient aggregation and bounded parameter drift without modifying model code or optimizers. The key insight is a two-stage defense against missing messages:  
(i) \emph{Unbiased gradient aggregation}—each worker reconstructs a consistent gradient estimate from whatever packets arrive, guaranteeing expectation-level correctness; and  
(ii) \emph{Bounded-drift parameter broadcasts}—we prove the inter-worker model discrepancy remains \(O(1)\) even after arbitrarily many iterations, preventing the unbounded divergence typical of asynchronous setups. Analytical bounds are matched by experiments on LLAMA2 7B model with 64 GPUs: tolerating \(10\%\) random packet loss yields \(\le 0.8\%\) perplexity change.
This work bridges the gap between communication-efficient datacenter protocols and the accuracy and generalization guarantees demanded by modern large-model training, enabling robust, high-throughput learning on commodity or wide-area networks.
\end{abstract}

\section{Introduction}

Distributed training of large neural network models has rapidly become essential due to increasing model sizes and data volumes. State-of-the-art language models, for instance, often require scaling across hundreds or thousands of GPUs \citep{tithi2025scaling, epoch2025trends}. While parallelization strategies such as data, model, and pipeline parallelism combined with advanced optimizers (e.g., ZeRO-2) have addressed many scalability issues, communication overhead and synchronization remain critical bottlenecks \citep{jin2025megascalemoe}. Conventionally, distributed training relies on reliable connection (RC) communication protocols that guarantee reliable packet delivery, resulting in significant overhead and high tail latencies as clusters scale \citep{gangidi2024roce}. \\
\\
Recent research has started exploring relaxed reliability assumptions, leveraging loss-tolerant communication protocols to mitigate synchronization delays \citep{wang2024mlt}. Such approaches demonstrate potential gains in training throughput, albeit at the risk of gradient or parameter inconsistencies due to packet loss. While gradient compression techniques significantly reduce bandwidth requirements \citep{jia2024sdp4bit}, they typically presume reliable delivery; packet drops in these scenarios can severely degrade performance. Similarly, techniques that introduce bounded asynchronous, such as Stale Synchronous Parallel (SSP), relax synchronization constraints to manage slow workers and network jitter, yet they still rely fundamentally on consistent communication \citep{kim2025halos}. \\ 
\\
A key challenge remains largely unaddressed: enabling efficient, large-scale distributed training under genuinely unreliable networking conditions without compromising convergence or final model accuracy. Specifically, quantifying the effects of unreliable packet delivery on model divergence and devising principled, robust compensation strategies are critical open issues. \\ 
\\
In this work, we address these gaps by developing a novel framework for scalable distributed training explicitly designed for unreliable networks. We introduce unbiased gradient aggregation methods resilient to packet loss, analytically quantify the resulting model divergence, and propose practical strategies to maintain drift within acceptable bounds. By doing so, we demonstrate that it is possible to leverage the throughput advantages of loss-tolerant protocols while still achieving stable convergence and high accuracy. Our analytical and empirical findings offer a path forward for robust distributed training under realistic network conditions, bridging an important gap in current large-scale learning frameworks.

\section{Previous Work}

Distributed training methods have advanced significantly to address scalability and communication bottlenecks inherent in training large neural network models. Optimizers such as ZeRO-2 partition model states across GPUs and overlap computation with communication, reducing memory consumption and enabling the efficient scaling of extremely large models \citep{rajbhandari2019zero}. Despite these innovations, traditional distributed training typically employs TCP for reliable communication, resulting in significant synchronization overhead and increased latency, particularly as cluster sizes expand.\\
\\
To mitigate these challenges, recent studies have explored using loss-tolerant communication protocols over UDP, accepting controlled packet loss to improve throughput. For example, the Loss-tolerant Transmission Protocol (LTP) was introduced to reduce synchronization delays by tolerating packet losses without significantly harming convergence \citep{chen2023boosting}. Similarly, Differential Gradient Transmission (DGT) adopts a selective gradient-update strategy based on contributions to model accuracy, thereby providing robustness even under unreliable communication conditions \citep{yang2023DGT}.\\
\\
Alongside protocol-level innovations, gradient compression techniques have significantly reduced the bandwidth requirements of distributed training. Methods such as Deep Gradient Compression (DGC) combine momentum correction, gradient clipping, and sparsification, achieving substantial reductions in communication overhead without compromising accuracy \citep{lin2017deep}. Other approaches, including Top-$k$ sparsification, quantization methods like QSGD, and low-rank approximations such as PowerSGD, further optimize the balance between communication efficiency and convergence quality \citep{evaluation2023gradient}. However, most gradient compression techniques implicitly assume reliable gradient delivery, potentially limiting their effectiveness under packet loss scenarios.\\
\\
Moreover, synchronization strategies have progressively relaxed strict synchronous constraints. Stale Synchronous Parallel (SSP) methods introduce bounded asynchronous updates, accommodating slow workers and network jitter \citep{ho2013lazytable}. Dynamic SSP further adapts the staleness threshold based on runtime conditions to improve training efficiency while still fundamentally depending on consistent network communication \citep{zhao2019dssp}.\\
\\
Finally, efficient implementations of collective communication primitives, such as All-Reduce using ring or butterfly algorithms, have become essential to achieving contention-free, bandwidth-optimal parameter aggregation \citep{patarasuk2009bandwidth,wiki_collective}.\\
\\
Despite these advances, current approaches either compromise accuracy to handle packet loss or strictly enforce synchronization, restricting scalability under genuinely unreliable network conditions. The present work uniquely addresses this gap by developing a robust analytical and empirical framework designed explicitly for training stability and accuracy under lossy communication scenarios.

\section{Model Formulation}
\label{sec:model}

We train a model with parameter vector \(\theta\in\mathbb{R}^d\) split into \(N\) shards
\[
\theta = \bigl(\theta^{(1)}, \dots, \theta^{(N)}\bigr),
\]
where shard \(\theta^{(j)}\) is owned by worker \(W_j\).

\paragraph{(1) Local gradient computation.}
At iteration \(t\), each worker \(W_i\) samples a mini-batch \(\mathcal{B}_t^{(i)}\) and computes its local gradient on all shards:
\[
g_t^{(i)} = \nabla_\theta L\bigl(\theta_t^{(i)};\,\mathcal{B}_t^{(i)}\bigr),
\]
which we partition into \(N\) pieces
\(
g_t^{(i)} = \bigl(g_t^{(i,1)},\dots,g_t^{(i,N)}\bigr)
\),
where \(g_t^{(i,j)}\in\mathbb{R}^{|\theta^{(j)}|}\).

\paragraph{(2) Gradient communication}
Workers exchange their gradient pieces over UDP.  A transmission of \(g_t^{(i,j)}\) from \(W_i\) to \(W_j\) succeeds with probability \(1-p\).  Denote the indicator
\(
s_t^{(i,j)}\sim\mathrm{Bernoulli}(1-p).
\)
Each \(W_j\) aggregates whatever shards it receives into an unbiased estimate:
\[
\hat g_t^{(j)} 
= \frac{\sum_{i=1}^N s_t^{(i,j)}\,g_t^{(i,j)}}
       {\sum_{i=1}^N s_t^{(i,j)}}
\quad\bigl(\mathbb{E}[\hat g_t^{(j)}]=G_t^{(j)}\bigr).
\]

\paragraph{(3) Local update.}
Worker \(W_j\) updates only its shard:
\[
\theta_{t+1}^{(j)}
= \theta_t^{(j)} - \eta_t\,\hat g_t^{(j)}.
\]

\paragraph{(4) Parameter broadcast.}
Immediately after updating, \(W_j\) broadcasts \(\theta_{t+1}^{(j)}\) to all other workers via unreliable connection.  Let
\(
r_t^{(j,i)}\sim\mathrm{Bernoulli}(1-p)
\)
indicate a successful reception by \(W_i\).  Each \(W_i\) then sets
\[
\theta_{t+1}^{(i,j)} =
\begin{cases}
\theta_{t+1}^{(j)}, & r_t^{(j,i)} = 1,\\
\theta_t^{(i,j)},   & r_t^{(j,i)} = 0,
\end{cases}
\]
so that \(W_i\) holds a potentially stale full model
\(\theta_{t+1}^{(i)}=(\theta_{t+1}^{(i,1)},\dots,\theta_{t+1}^{(i,N)})\).

\subsection*{Quantifying Model Drift}

Packet drops in parameter broadcasts introduce discrepancies between replicas of the same shard.  We now formalize and bound this “drift.” Unlike conventional distributed training—where missed gradient updates can lead to unbounded divergence between models over time (i.e., a growing consensus problem)—our approach ensures that the expected distance between any two models remains bounded with time \(t\). Specifically, we show that this discrepancy is \(O(1)\), rather than increasing with \(t\), due to the periodic resetting effect of successful broadcasts.

\begin{theorem}[Bounded Model Drift]
Under independent packet‐drop probability \(p<1\), the steady‐state expected squared discrepancy
\[
D_t = \theta_t^{(i,j)} - \theta_t^{(k,j)}
\]
between any two workers \(W_i,W_k\) for shard \(j\) satisfies
\[
\lim_{t\to\infty}\mathbb{E}[D_t^2]
= \frac{2p}{1+p}\,\sigma^2
= O(1).
\]
\end{theorem}

\paragraph{Proof.}
Let \(\Delta\theta_t = \theta_{t+1}^{(j)} - \theta_t^{(j)}\) and denote
\[
D_t = \theta_t^{(i,j)} - \theta_t^{(k,j)}.
\]
After a broadcast, each indicator \(s_t^{(i,j)},s_t^{(k,j)}\sim\mathrm{Bernoulli}(1-p)\) yields
\[
D_{t+1} = 
\begin{cases}
0, & s_t^{(i,j)}=s_t^{(k,j)}=1,\\
+\Delta\theta_t, & s_t^{(i,j)}=1,\;s_t^{(k,j)}=0,\\
-\Delta\theta_t, & s_t^{(i,j)}=0,\;s_t^{(k,j)}=1,\\
D_t, & s_t^{(i,j)}=s_t^{(k,j)}=0.
\end{cases}
\]
Squaring and taking expectations gives
\[
\mathbb{E}[D_{t+1}^2]
= (1-p)^2\cdot0
+2p(1-p)\,\mathbb{E}[\Delta\theta_t^2]
+p^2\,\mathbb{E}[D_t^2].
\]
Set \(E_t=\mathbb{E}[D_t^2]\) and use \(\mathbb{E}[\Delta\theta_t^2]=\sigma^2\).  Then
\[
E_{t+1} = p^2E_t + 2p(1-p)\,\sigma^2.
\]
Unrolling,
\[
E_t = (p^2)^tE_0 + 2p(1-p)\,\sigma^2\sum_{m=0}^{t-1}(p^2)^m
    = (p^2)^tE_0 + 2p(1-p)\,\sigma^2\frac{1-(p^2)^t}{1-p^2}.
\]
As \(t\to\infty\), \((p^2)^tE_0\to0\), so
\[
\lim_{t\to\infty}E_t
= \frac{2p(1-p)}{1-p^2}\,\sigma^2
= \frac{2p}{1+p}\,\sigma^2,
\]
establishing the \(O(1)\) drift bound. $\blacksquare$

\subsection*{Unbiased Gradient Aggregation}

We now show that, despite packet loss, the aggregated gradient remains unbiased and thus does not impair convergence in expectation.

Let \(g_i^{(j)}\) be the gradient computed by worker \(W_i\) on shard \(j\), and let
\[
S_{ij} \sim \mathrm{Bernoulli}(1-p)
\]
indicate successful delivery of \(g_i^{(j)}\) to the aggregation point at worker \(W_j\).  We form the resilient aggregate
\[
G_j \;=\; \frac{\sum_{i=1}^N S_{ij}\,g_i^{(j)}}{\sum_{i=1}^N S_{ij}},
\]
whenever \(\sum_iS_{ij}>0\) (and fall back to a small regularizer otherwise).

Since each \(g_i^{(j)}\) satisfies
\[
\mathbb{E}\bigl[g_i^{(j)}\bigr]
= G_j^*,
\]
the true full‐batch gradient, we condition on the random indicators \(\{S_{ij}\}\):
\[
\mathbb{E}\bigl[G_j \,\bigm|\,\{S_{ij}\}\bigr]
= \frac{\sum_{i=1}^N S_{ij}\,\mathbb{E}[g_i^{(j)}]}{\sum_{i=1}^N S_{ij}}
= \frac{\sum_{i=1}^N S_{ij}\,G_j^*}{\sum_{i=1}^N S_{ij}}
= G_j^*.
\]
Taking the full expectation then yields
\[
\mathbb{E}\bigl[G_j\bigr]
= G_j^*.
\]

\begin{corollary}
Even under independent packet‐drop probability \(p\), the aggregated shard gradient \(G_j\) remains an unbiased estimator of the true gradient \(G_j^*\).  Consequently, standard convergence guarantees for stochastic gradient methods continue to hold in expectation.
\end{corollary}

\subsection{System Preliminaries}
Dropping packets translates to missing gradients, activations, or parameters,
depending on the type of parallelism and collective being used. We assume a standard GPU cluster connected via RC stack (eg., IB, RoCE). Models are parallelised using a hybrid of data (DP), tensor (TP), and pipeline parallelism (PP), whose characteristics are reviewed succinctly in the Appendix, orchestrated by ZeRO-2 optimization scheme (see Algorithm 1): 
\[
\text{\tcbox[mylabel]{forward pass}}
\;\ra\;
\text{\tcbox[mylabel]{backward pass}}
\;\ra\;
\text{\tcbox[mylabel]{gradient synchronization (reduce-scatter)}}
\]
\[
\;\ra\;
\text{\tcbox[mylabel]{optimizer updates parameters}}
\;\ra\;
\text{\tcbox[mylabel]{parameter synchronization (all-gather)}}
\]
\\
\\During gradient and parameter synchronization steps, we simulate packet drops by randomly omitting shard updates between workers according to a predefined drop probability. Specifically, each parameter or gradient shard transmission from worker $W_i$ to $W_j$ succeeds with probability $(1 - p)$, and otherwise fails, resulting in each worker retaining stale values for lost shards. Despite these packet drops, our framework ensures unbiased gradient aggregation and bounded parameter drift, preserving convergence and accuracy guarantees. We use Megatron-LM, NVIDIA’s open-source research codebase for training very large Transformer-style language models, for packet-loss simulation \citep{weintraub2025simcode}.\\

\begin{algorithm}[!htbp]
  \caption{Drop simulation on worker $n$ at iteration $i$}
  \label{alg:loss-aware}
  \begin{algorithmic}[1]
    \Require model parameters $\theta_{n}$; micro-batches $\{\mathcal{B}^{(m)}_{n}\}_{m=1}^{M}$;
            loss rates $p_{\text{grad}},p_{\text{param}}$
    \Ensure  updated parameters $\theta_{n}$ (post broadcast)

    \For{$m=1$ \textbf{to} $M$} 
        \State $g^{(m)}_{n}\gets\nabla_{\theta}L\!\left(\mathcal{B}^{(m)}_{n},\theta_{n}\right)$ \Comment{gradient compute}
    \EndFor

    \State \textbf{reduce-scatter}$(g_{n})\to\{g^{(j)}_{n}\}$ \Comment{gradient sync}
    \ForAll{shards $j=1,\dots,N$ \textbf{in parallel}}
        \If{$\text{Uniform}(0,1)<p_{\text{grad}}$}
            \State \textbf{drop} gradient: $g^{(j)}_{n}\gets g^{\text{prev},(j)}_{n}$
        \Else
            \State $g^{\text{prev},(j)}_{n}\gets g^{(j)}_{n}$
        \EndIf
    \EndFor
    \State renormalise surviving shards to obtain $\hat g_{n}$

    \State \textbf{optimizer update}$(\theta^{(n)}_{\text{local}},\hat g_{n})$ \Comment{optimizer update}

    \State \textbf{all-gather}$(\theta^{(n)}_{\text{local}})\to\{\theta^{(j)}\}$ \Comment{parameters sync}
    \ForAll{incoming shards $j=1,\dots,N$ \textbf{in parallel}}
        \If{$\text{Uniform}(0,1)<p_{\text{param}}$}
            \State \textbf{drop} update: $\theta^{(j)}\gets\theta^{\text{prev},(j)}$
        \Else
            \State $\theta^{\text{prev},(j)}\gets\theta^{(j)}$
        \EndIf
    \EndFor
  \end{algorithmic}
\end{algorithm}

\section{Experiments}

We carried out our experiments on a cluster of two Gaudi\,4 nodes (64 Gaudi\,3 accelerators in total), training the LLaMA-2 7B model for 5\,000 iterations with the Megatron-LM framework, and evaluated four packet-drop settings: 10\%, 20\%, 30\%, and 40\% during the communication phases.
\\
\\

\begin{table}[h]
\centering
\small        
\setlength{\tabcolsep}{4pt} 
\begin{tabular}{lccccc}
\toprule
\textbf{Metric} & \textbf{0 \%} & \textbf{10 \%} & \textbf{20 \%} & \textbf{30 \%} & \textbf{40 \%} \\
\midrule
Train Loss   & 1.645 & 1.653 (+0.47\%) & 1.672 (+1.63\%) & 1.684 (+2.36\%) & 1.710 (+3.95\%) \\
Train PPL    & 5.183 & 5.223 (+0.77\%) & 5.325 (+2.72\%) & 5.388 (+3.95\%) & 5.531 (+6.71\%) \\
Val Loss  & 2.366 & 2.378 (+0.49\%) & 2.392 (+1.10\%) & 2.403 (+1.56\%) & 2.430 (+2.72\%) \\
Val PPL   & 10.653 & 10.778 (+1.17\%) & 10.934 (+2.63\%) & 11.055 (+3.77\%) & 11.362 (+6.65\%) \\
\bottomrule
\end{tabular}
\captionsetup{position=bottom}
\caption{LLAMA-2 7B (5K iterations). Each non-baseline entry shows the raw value followed by its relative change compared with the 0\% packet-loss run.}
\label{tab:llama-packet-loss-4}
\end{table}

Our experiments shows training tolerates up to $\sim20\%$ random shard loss with $<\!3\%$ degradation in accuracy, while convergence and generalization remain virtually intact.  At $30$--$40\%$ the model still trains, but quality begins to erode more noticeably.\\

\begin{figure}[H]
  \centering
  \includegraphics[width=0.9\textwidth]{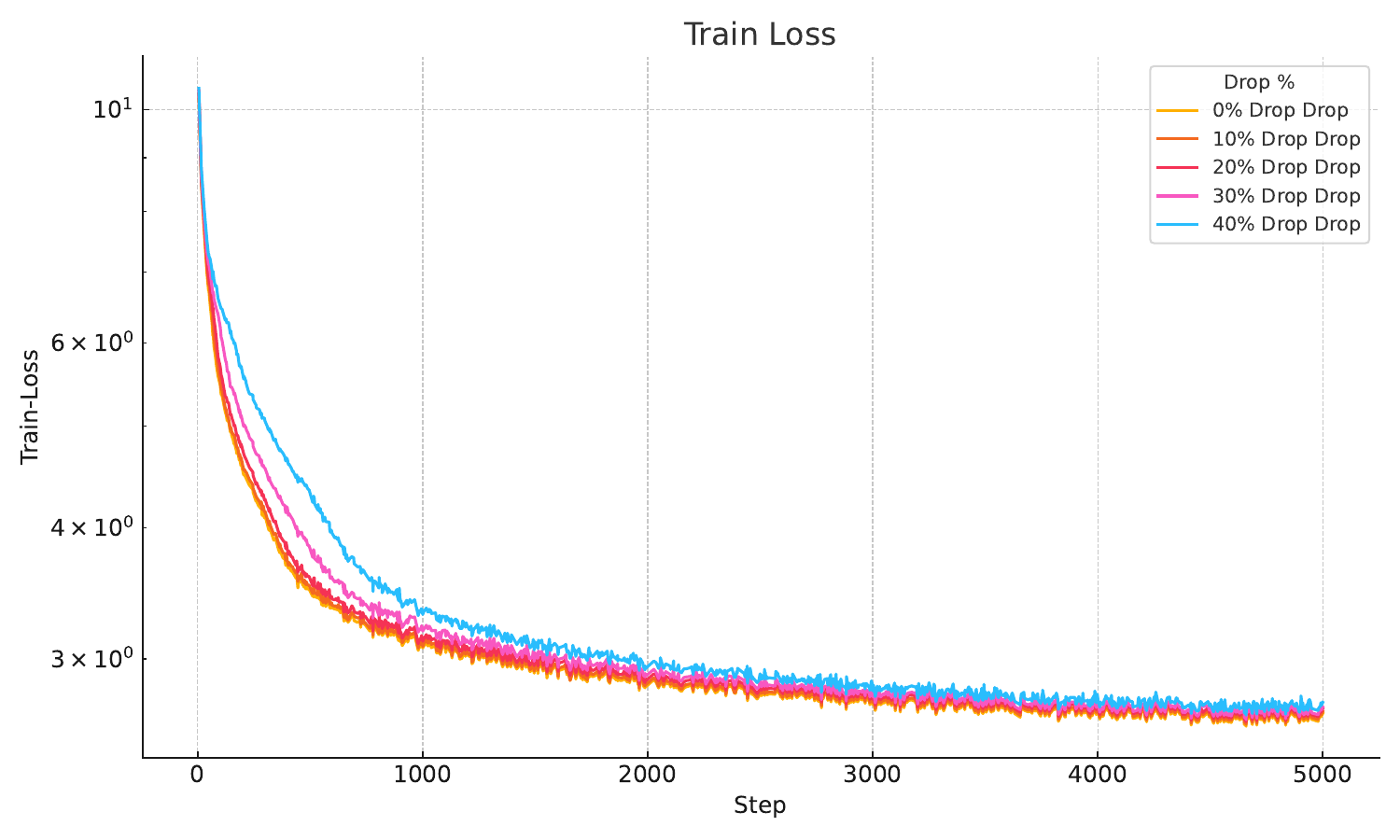}
  \caption{Train‑loss (logarithmic scale) vs.\ training steps for LLAMA‑2‑7B at five drop rates (0–40\%). Colors correspond to the percentages indicated in the legend.}
  \label{fig:llama2-drop-all}
\end{figure}

\section{Discussion}
\textbf{Summary.}
This work demonstrates, both analytically and empirically, that large-scale neural network training can remain stable even when the underlying network delivers gradients and parameters over \emph{unreliable} transport.  
Our two–stage defense—unbiased gradient aggregation and bounded-drift parameter broadcasts—guarantees expectation-level correctness while provably capping inter-worker model divergence at \(O(1)\), independent of training length.  
Experiments on LLAMA2-7B (64~GPUs) showed that tolerating up to \(10\%\) random packet loss changes validation perplexity by at most \(0.8\%\), all while eliminating the acknowledgment traffic and retransmission storms that plague traditional reliable connections at scale.  
These results bridge the long-standing gap between high-throughput datacenter transport protocols and the convergence guarantees required by modern language and vision models.\\
\\
\textbf{Limitations.}
Our theory assumes i.i.d.\ Bernoulli packet drops with a fixed rate \(p\).  Real networks exhibit bursty and correlated loss patterns; extending the analysis to such regimes is an open question. We evaluated on clusters with identical GPUs and bandwidth; heterogeneous environments may aggravate straggler effects and interact non-trivially with our bounded-drift guarantees. While our method is orthogonal to sparsification and quantization, we did not measure combined effects; lossy delivery could amplify compression bias. Each worker must track per-iteration reception masks and perform local renormalization. For very small tensors the extra computation can dominate communication savings. Benchmarks focused on autoregressive language modeling; vision transformers and reinforcement-learning workloads, which have different gradient statistics, remain to be validated. \\
\\
\textbf{Future Directions.}
Utilize hybrid reliable–unreliable transport. Future systems could dynamically classify traffic by \emph{importance}: large-norm gradients and rare vocabulary embeddings would traverse a reliable channel (RC), while the long-tail of small-magnitude updates would continue over a best-effort UDP stream. Such selective reliability promises near-optimal throughput with minimal accuracy sacrifice.\\
Erasure coding or lightweight parity packets can be injected only for shards whose loss would disproportionately slow convergence. Coupling this with our unbiased aggregation could shrink the effective loss rate for critical updates without reinstating full TCP-style handshakes.\\ 
Packet-drop tolerance is presently a static hyper-parameter.  
An interesting avenue is to make \(p\) adaptive—monitoring gradient variance and automatically tightening reliability as training nears convergence, akin to learning-rate schedules. \\
Combining our framework with Top-\(k\) sparsification or PowerSGD may compound bandwidth savings, but requires revisiting bias correction under simultaneous quantization \emph{and} random loss.\\
To guarantee reproducibility and simplify debugging on multi-thousand-GPU
systems, future versions should log every shard-level routing decision—
specifically, whether each worker retained the previously received value or
applied an updated one—so that the entire training step sequence can be
re-played deterministically.\\
Explore if reducing lr or adding regularisation could recover part of that gap even at higher packet-loss levels.\\
Finally, implementing unbiased aggregation and drift tracking in the NIC or smart switch could reduce CPU overhead and enable wire-speed operation even at 400 Gb s\(^{-1}\) links. \\
Taken together, these directions aim to evolve the proposed framework from a proof-of-concept into a drop-in communication layer for the next generation of trillion-parameter models.

\section{Acknowledgments}
We thank Niv Giladi for technical advising and valuable comments on the manuscript.

\clearpage
\begin{appendices}
\renewcommand{\thesection}{Appendix \Alph{section}}
\section{Collective Communication Primitives}
  
  \textbf{Communication collectives.} Collective communication primitives are building blocks for interaction patterns between processes. In distributed training, collective communication is handy for exchanging parameters between computation nodes. The collective primitive being used the most is \textit{All-Reduce}.\\
  \textbf{\textit{All-Reduce}.} Given $N$ computing nodes $w_i$, $i\in\{1,2,\dots,N\}$ and parameters $\vect{x}_i$ on node $w_i$, \textit{All-Reduce} aggregates the parameters by an operator $\otimes$, usually SUM, and distributes the result to all computing nodes, such that each node holds $\vect{x}_1\otimes \vect{x}_2\otimes\dots \vect{x}_N$. \textit{All-Reduce} can be interpreted as a \textit{Reduce} operation with a subsequent \textit{Broadcast}, as implemented by the centralized parameter-server paradigm. \textit{All-Reduce} is also equivalent to a \textit{Reduce-Scatter} operation with a subsequent \textit{All-Gather}. This method is decentralized, yet synchronous, and provides better scalability properties.\\
  \textbf{\textit{Reduce-Scatter}.} Given $N$ computing nodes $w_i$, $i\in\{1,2,\dots,N\}$ and parameters $\vect{x}_{i}^{1}\oplus \vect{x}_{i}^{2}\oplus\dots \vect{x}_{i}^{N}$ ($\oplus$ denotes concatenation) on node $i$, \textit{Reduce-Scatter} aggregates the parameters by an operator $\otimes$, usually SUM, and store each subset of the parameters on a different node. Such that node $i$ holds $\vect{x}_{1}^{i}\otimes \vect{x}_{2}^{i}\otimes\dots \vect{x}_{N}^{i}$.\\
  \textbf{\textit{All-Gather}.} Given $N$ computing nodes $w_i$, $i\in\{1,2,\dots,N\}$ and parameters $\vect{x}_i$ on node $i$, \textit{All-Gather} concatenates the parameters and distributes the result to all computing nodes, such that each node holds $\vect{x}_1\oplus \vect{x}_2\oplus\dots \vect{x}_N$.\\
  These communication primitives do not specify the actual data flow, and for each such communication primitive there multiple implementations with different bounds on latency and communication cost.\\

\section{Train Loss}
\begin{figure}[H]
  \centering
  \begin{subfigure}{0.48\textwidth}
    \centering
    \includegraphics[width=\linewidth]{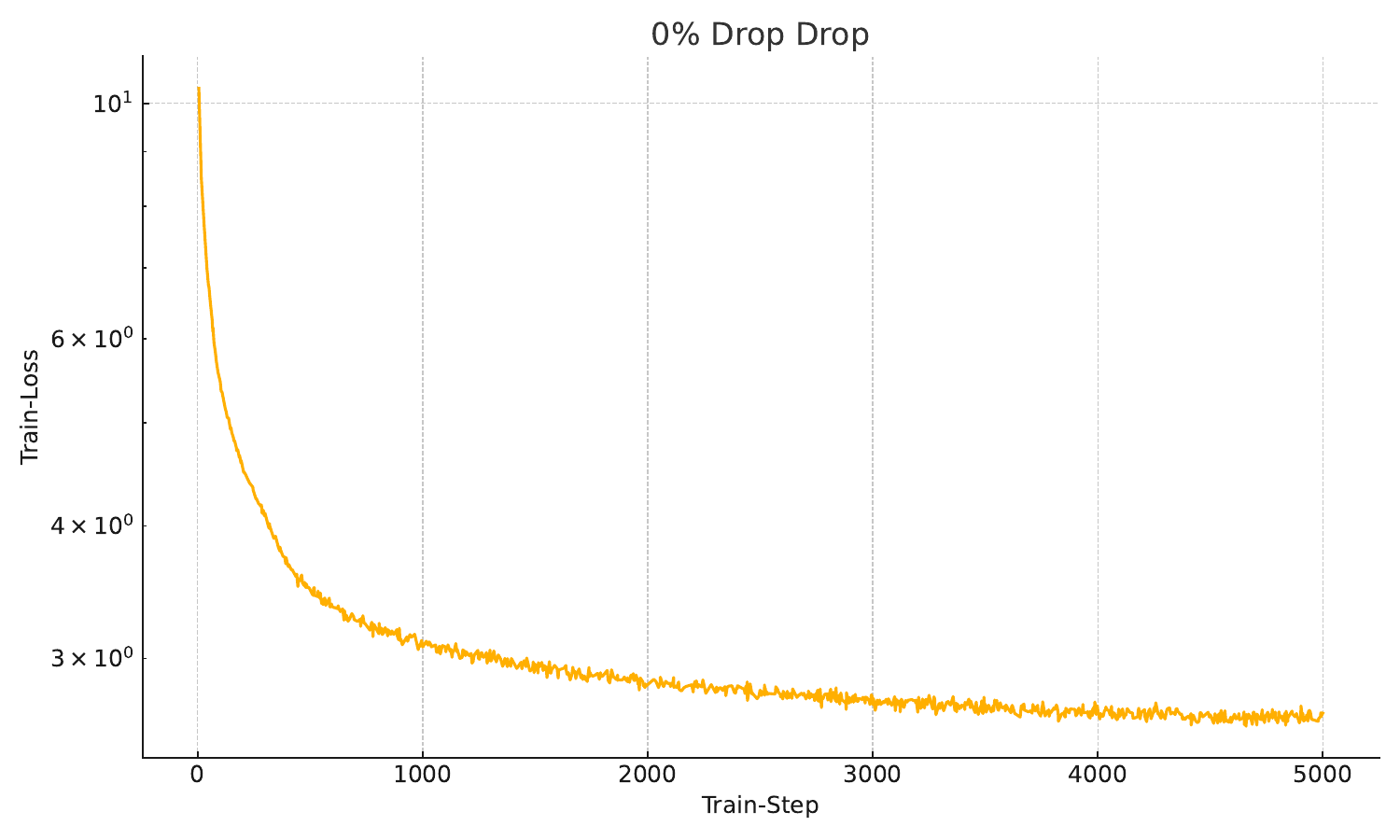}
    \caption{0\% Drop}
  \end{subfigure}
  \hfill
  \begin{subfigure}{0.48\textwidth}
    \centering
    \includegraphics[width=\linewidth]{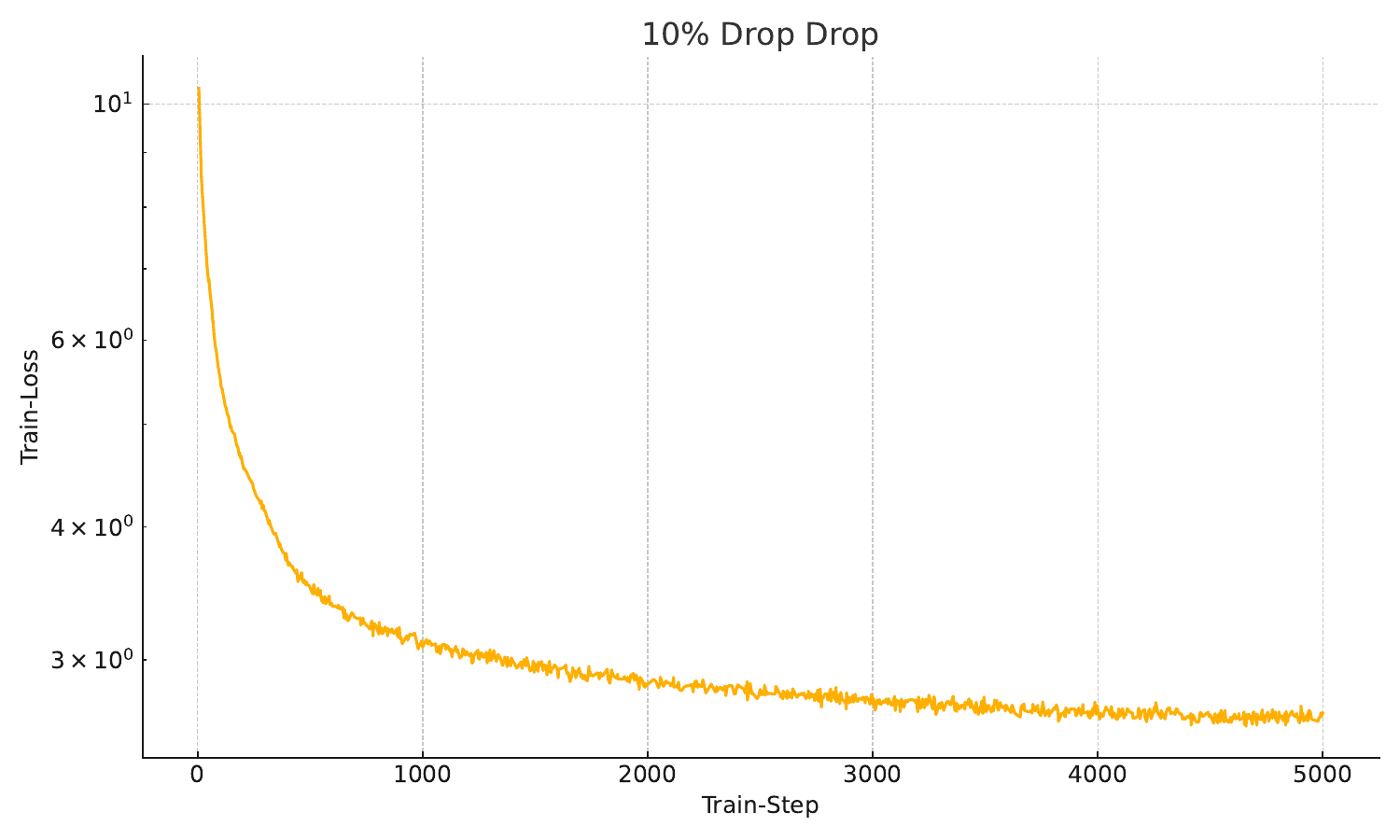}
    \caption{10\% Drop}
  \end{subfigure}

  \begin{subfigure}{0.48\textwidth}
    \centering
    \includegraphics[width=\linewidth]{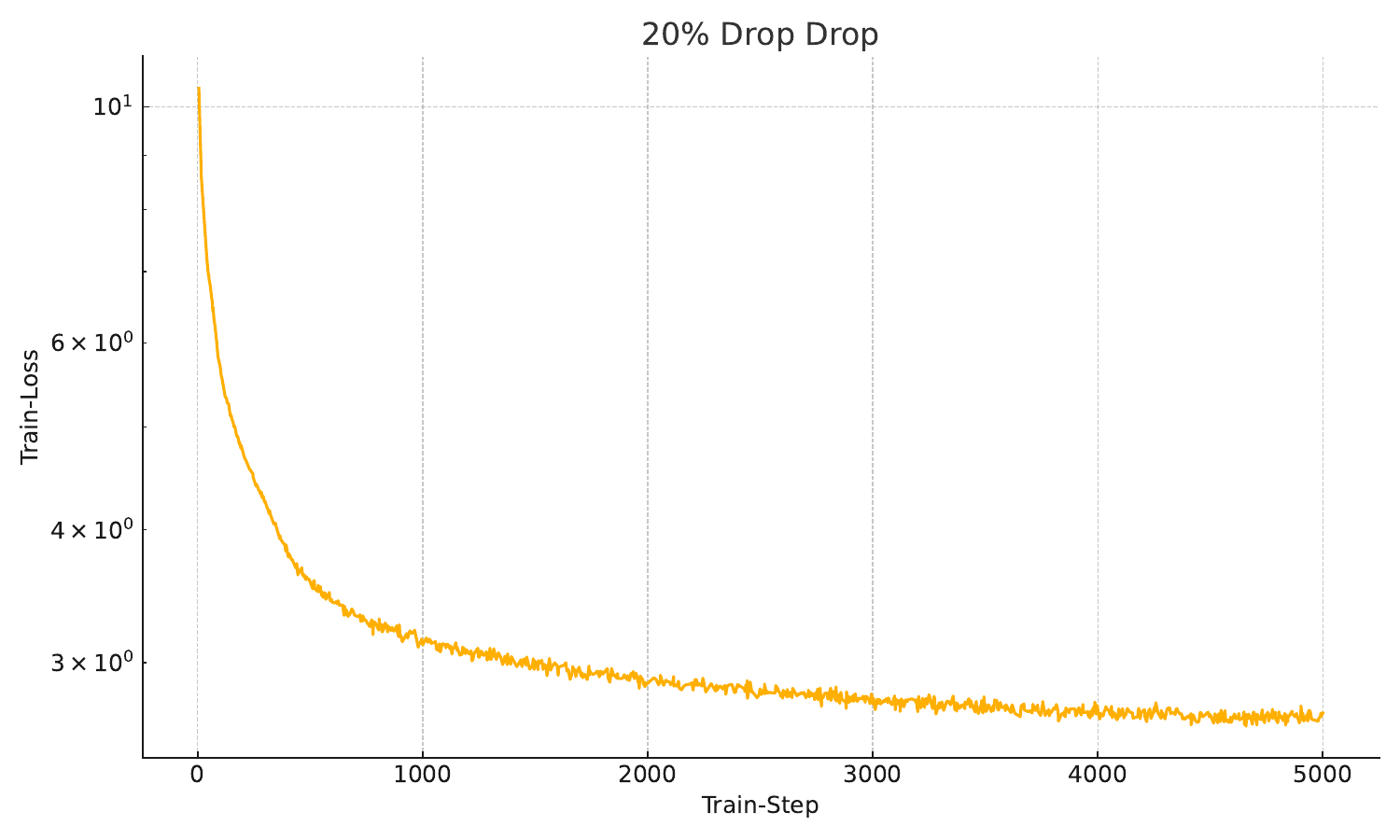}
    \caption{20\% Drop}
  \end{subfigure}
  \hfill
  \begin{subfigure}{0.48\textwidth}
    \centering
    \includegraphics[width=\linewidth]{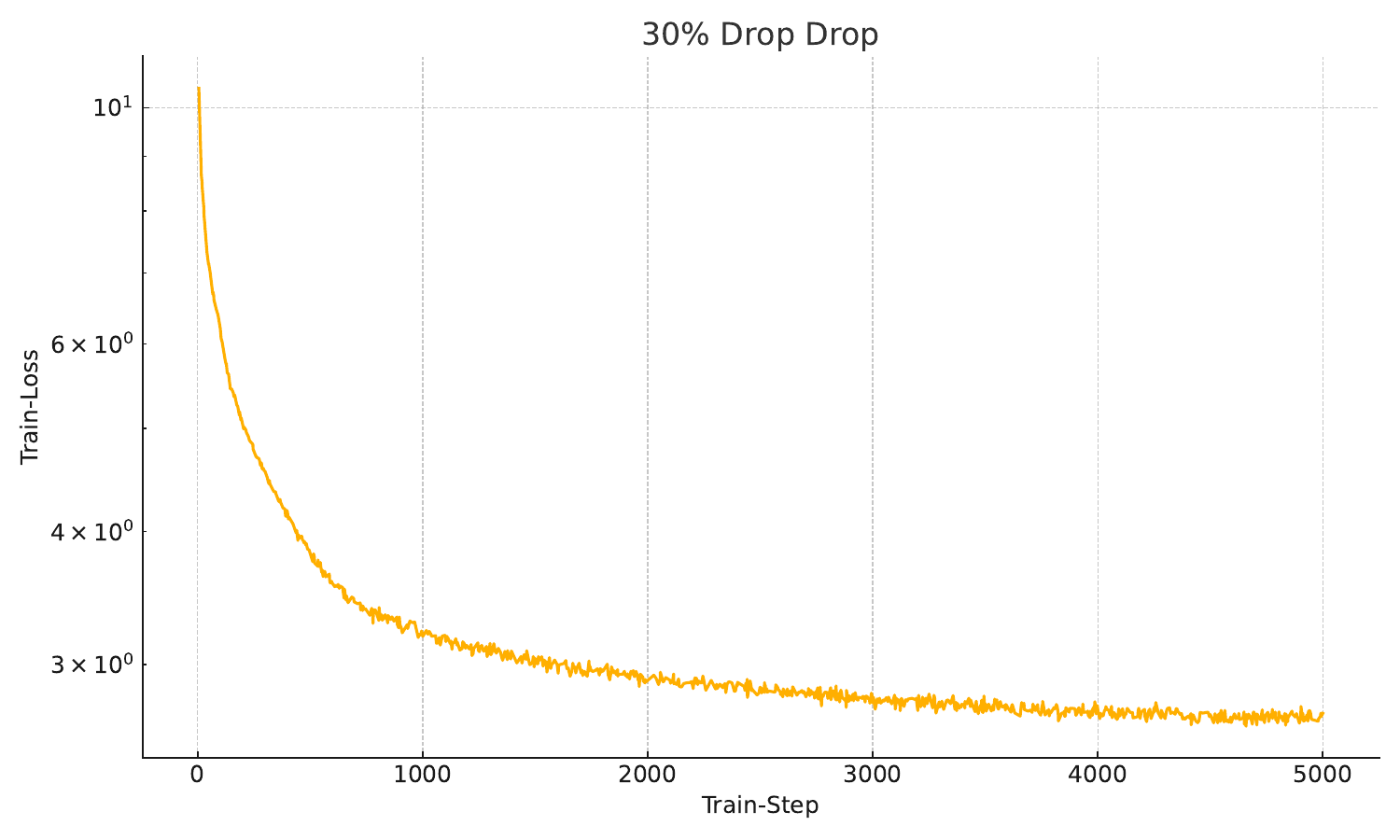}
    \caption{30\% Drop}
  \end{subfigure}

  \begin{subfigure}{0.48\textwidth}
    \centering
    \includegraphics[width=\linewidth]{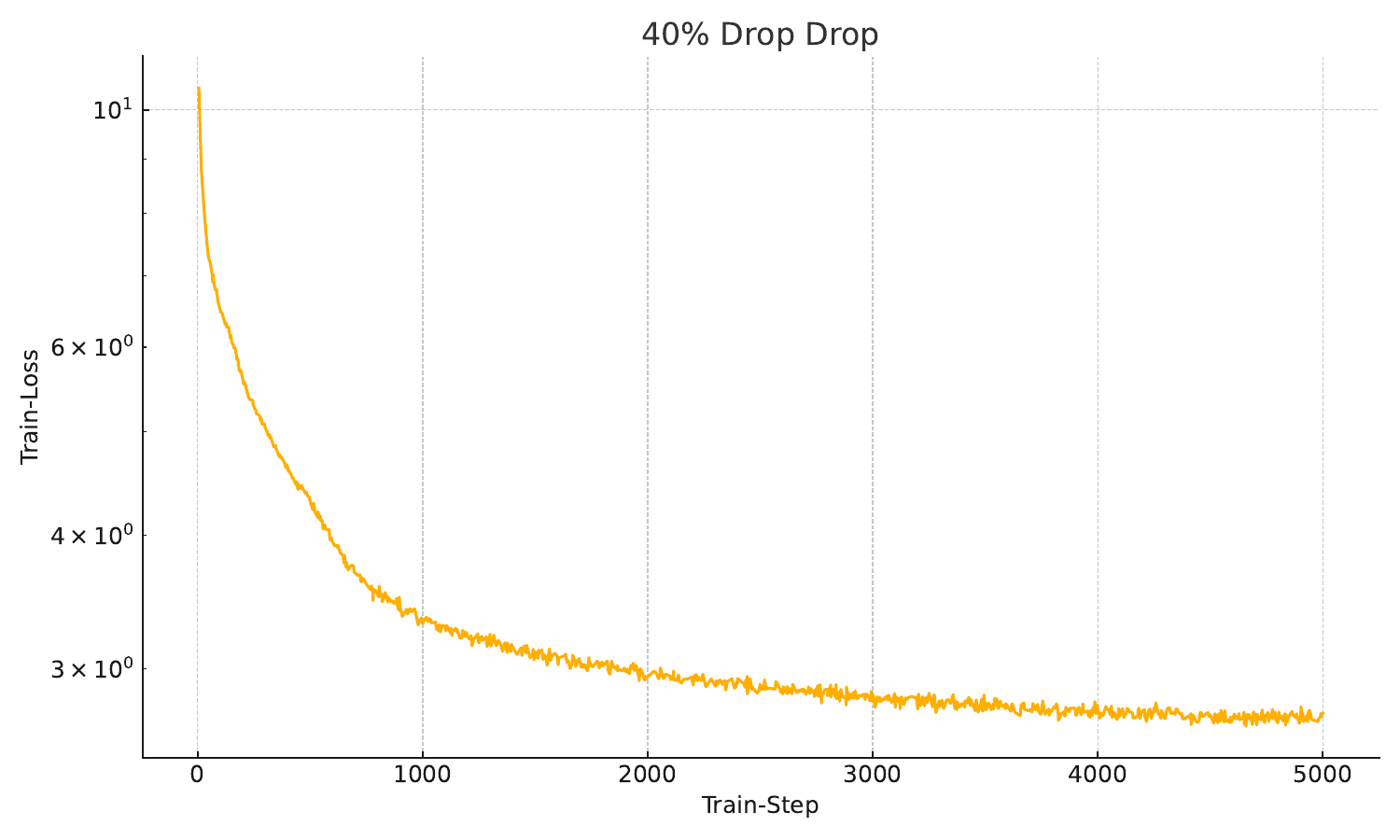}
    \caption{40\% Drop}
  \end{subfigure}

  \caption{Training loss (logarithmic scale) vs.\ training steps for LLAMA‑2‑7B under increasing drop rates.}
  \label{fig:llama2-drop-ablation}
\end{figure}

\clearpage

\end{appendices}

\bibliographystyle{plain}
\bibliography{references}

\end{document}